\documentclass[aps,twocolumn,prl,floatfix,showpacs,lengthcheck,citeautoscript,superscriptaddress,nobibnotes,longbibliography]{revtex4-2}
\usepackage{amsmath}
\usepackage{amssymb}
\usepackage{graphicx}
\usepackage{bm}
\usepackage{color,soul}
\usepackage{braket}
\usepackage[dvipsnames,svgnames,table]{xcolor}
\usepackage{times}
\usepackage{MnSymbol}

\usepackage{bbold}
\usepackage{hyperref}
\usepackage{orcidlink}
\usepackage{lipsum}

\definecolor{NathanBlue}{rgb}{0.,0.24,0.51}
\newcommand{\Blue}{\color{NathanBlue}}

\newcommand{\Heff}{H_\mathrm{eff}}
\newcommand{\Hsyk}{H_\mathrm{SYK}}
\newcommand{\Hkdbh}{H_\mathrm{KDBH}}
\newcommand{\SFF}{\mathrm{SFF}}

\hypersetup{
pdfstartview={FitH},% fits the width of the page to the window
colorlinks=true,    % false: boxed links; true: colored links
linkcolor=NavyBlue, % color of internal links
citecolor=NavyBlue,   % color of links to bibliography
filecolor=NavyBlue, % color of file links
urlcolor=NavyBlue   % color of external links
}

\begin{document}
\title{{\Blue Sachdev-Ye-Kitaev physics from the Hubbard model:
A Floquet engineering approach}}
%%%%%%%%%%%%%%%%%%%%%%%%%%%%%%%%%%%%%%%%%%
\author{Charles Creffield
\orcidlink{0000-0003-1044-9218}}
\email{c.creffield@fis.ucm.es}
\affiliation{Departamento de F\'isica de Materiales, Universidad
Complutense de Madrid, E-28040 Madrid, Spain}
\author{Fernando Sols
\orcidlink{0000-0002-0947-286X}}
\affiliation{Departamento de F\'isica de Materiales, Universidad
Complutense de Madrid, E-28040 Madrid, Spain}
\author{Marco Schir\`o}
\affiliation{JEIP, UAR 3573 CNRS, Coll\`{e}ge de France, PSL Research University, 11 Place Marcelin Berthelot, 75321 Paris Cedex 05, France}
%%%%%%%%%%%%%%%%%%%%%%%%%%%%%%%%%%%%%%%%%%
\author{Nathan Goldman\,
\orcidlink{0000-0002-0757-7289}}
\email{nathan.goldman@lkb.ens.fr}
\affiliation{Laboratoire Kastler Brossel, Coll\`ege de France, CNRS, ENS-Universit\'e PSL, Sorbonne Universit\'e, 11 Place Marcelin Berthelot, 75005 Paris, France}
\affiliation{International Solvay Institutes, 1050 Brussels, Belgium}
%%%%%%%%%%%%%%%%%%%%%%%%%%%%%%%%%%%%%%%%%%
%%%%%%%%%%%%%%%%%%%%%%%%%%%%%%%%%%%%%%%%%%%%%%%%%%%%%%%%%%
\begin{abstract}
The Sachdev-Ye-Kitaev (SYK) model has attracted widespread attention
due to its relevance to diverse areas of physics, such as high temperature superconductivity,
black holes, and quantum chaos. The model is, however, extremely challenging to realize
experimentally. In this work, we show how a particular form of Floquet engineering, termed
``kinetic driving'', effectively eliminates single-particle processes and creates quasi-random all-to-all interactions when applied
to models of Hubbard type. For the specific case of the Bose-Hubbard model, 
we explicitly verify that the driven system indeed reproduces
SYK physics by direct comparison of the spectral form factor and out-of-time ordered
correlation functions (OTOCs). Our findings indicate that a cold-atom realization of kinetic driving -- achieved through modulation of hopping amplitudes in an optical lattice -- offers a practical and accurate platform for quantum simulation of the SYK model.
\end{abstract}
\date{\today}
\maketitle
%%%%%%%%%%%%%%%%%%%%%%%%%%%%%%%%%%%%%%%%%%%%%%5
\textit{Introduction ---}
The Sachdev-Ye-Kitaev (SYK) model \cite{sachdev_1992,kitaev} has become a topic of intense 
theoretical interest in recent years. The model describes random, all-to-all interactions between
fermions, where the particles move collectively in pairs with no single-particle hopping
processes. A consequence of its extremely dense low-energy spectrum is that
the SYK model lacks quasiparticle excitations, making it a natural candidate to describe
non-Fermi liquid physics. Its applications range \cite{sachdev_review} from modeling the strange 
metal phase in high temperature superconductors, to holographic quantum matter and information scrambling 
in black holes, and it has become a paradigmatic system for the study of extreme many-body quantum chaos. 
Its solvability in the large-$N$ limit, where $N$ is the number of fermion flavors, thus provides
an attractive way of exploring these phenomena in a tractable manner.
The bosonic version of the SYK model has received considerably less attention. While the bosonic representation is known to display different low-temperature physics, notably with a glassy phase with replica symmetry breaking, its high-energy properties share much of the features of its fermionic counterpart~\cite{georges2000mean,georges2001quantum,fu_sachdev,baldwin2020quenched,swingle2024bosonic,liu2025dissipative}.

Despite the model's simplicity from a theoretical point of view, 
the complicated long-range nature of the interactions
makes it extremely challenging to realize experimentally. Various suggestions have been made
for solid-state implementations such as a graphene flake with an irregularly shaped boundary \cite{graphene,kruchkov_2020,kruchkov_2023,kruchkov_2024},
or by using Majorana modes hosted by superconducting wires \cite{majorana_wires} or the surface of
a topological insulator \cite{majorana_chip}. The excellent controllability of ultracold atoms held
in optical lattice potentials also provides an appealing pathway towards simulating
the SYK model \cite{kagome, danshita}, with Rydberg arrays \cite{rydberg} and cavity-QED systems
\cite{hauke_cavityQED, hauke_simulation} offering similar advantages. Digital quantum circuits
have been proposed too \cite{digital_sim_1, digital_sim_2}, and small-scale implementations 
of SYK systems have been achieved recently on Google's Sycamore \cite{wormholes} and IBM
quantum processors \cite{noisy,IBM}.   

An alternative approach to developing complicated laboratory experiments
is to {\em engineer} the required interactions by using a Floquet approach,
and driving a relatively simple system. In the Floquet technique, a parameter of
the system is varied periodically, and in the high-frequency regime the system can
then be described by an effective static Hamiltonian, consisting of terms that have
been renormalized or created by the driving. In principle, these various terms can be calculated 
perturbatively through the inverse-frequency or Magnus expansions \cite{magnus,goldman2014periodically,bukov2015universal,eckardt2017colloquium}. A well-known
example of this type of engineering is the periodically-driven Bose-Hubbard model.
By oscillating the lattice potential  -- ``shaking'' the system --  the tunneling can be tuned to
zero, allowing the Mott transition to be dynamically controlled \cite{arimondo}.
More complicated driving schemes can be devised to produce complex tunneling elements~\cite{bermudez2011synthetic,kolovsky2011creating,kolovsky_comment,struck2012tunable,hauke2012non,goldman2015periodically,flux},
allowing driven systems to emulate synthetic gauge fields and topological band structures~\cite{goldman2016topological,cooper2019topological,weitenberg2021tailoring}. Beyond the control over single-particle properties, Floquet engineering can also be used to effectively create complex interaction processes, including multi-body interactions~\cite{ma2011photon,daley2014effective,hafezi2014engineering,lee2018floquet}, correlated hopping~\cite{rapp2012ultracold,di2014quantum,meinert2016floquet,gorg2019realization,barbiero2019coupling,schweizer2019floquet,kamal2024floquet}, pair hopping~\cite{zahn2022formation,goldman2023floquet,defossez2025dynamic}, spin-spin interactions~\cite{hung2016quantum,choi2020robust,geier2021floquet,sun2023engineering,nishad2023quantum}, and chiral p-wave pairing~\cite{dehghani2021light}. 

\begin{figure}
\begin{center}
\includegraphics[width=0.5\textwidth,clip=true]{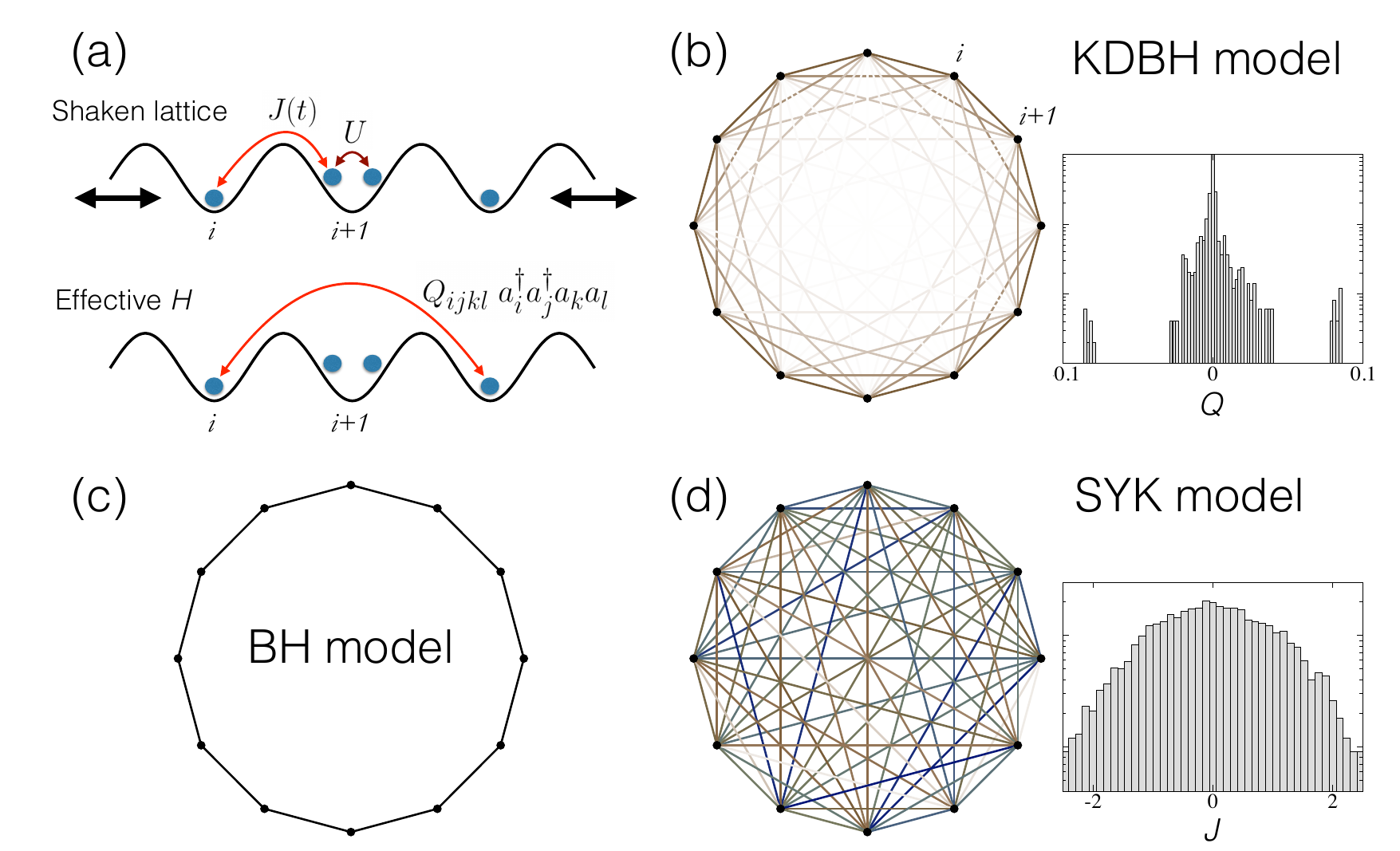}
\end{center}
\caption{(a) Particles held in an optical lattice can be well described by the Hubbard model
(Eq.~\eqref{hubbard}), with
nearest neighbor hopping $J$ and onsite repulsion $U$. Shaking the lattice allows us to
make $J$ a time periodic function At high driving frequencies this produces a static effective 
model with long-ranged
interaction terms of the form $Q_{i j k l} a^\dagger_i a^\dagger_j a_k a_l$ describing
processes like doublon hopping and assisted tunneling.
(b) Schematic representation of the interaction amplitudes of the KDBH model (Eq. \ref{effective_H})  shaded according 
to their magnitude, 
$\left| Q_{i j k l} \right|$. The strongest links are between nearest neighbors, and their
magnitude falls as the distance between the sites increases. Inset: The cumulative distribution of $Q_{i j k l}$ for $0.6 \leq \kappa \leq 0.9$, where $\kappa$ is sampled in steps of 0.01. The histogram is plotted on a logarithmic scale. The distribution is strongly peaked at small $Q$, and 
unlike the distribution of the SYK model, has gaps.
(c) In the Bose-Hubbard model a given site $i$ is only connected to its
nearest-neighbors by single-particle tunneling processes.
(d) As in (b) but for the SYK model (Eq. \ref{H_SYK}). Every site is connected to every other
site by a random tunneling amplitude.
Inset: In contrast to the effective model, the distribution of the amplitudes
$J_{i j; k l}$ is dense and is of Gaussian form, so that when plotted logarithmically
the histogram is parabolic.}
\label{schematic_plot}
\end{figure}

In this work, we employ a driving protocol known as ``kinetic driving" \cite{kinetic_epl,kinetic_driving_2019,kinetic_driving_2023}, in which the kinetic energy term of a system is periodically modulated such that its time-averaged value vanishes. When this driving is applied to a Hubbard-type lattice model, the resulting effective Hamiltonian simplifies to a sum of all-to-all quartic interaction terms of the form $Q_{ijkl}a_i^\dagger a_j^\dagger a_k a_l$, characterized by a broad distribution of coupling strengths -- closely resembling the structure of the SYK model; see Fig.~\ref{schematic_plot}(a). As the general form of the effective interactions does not depend on 
the statistics of the particles, our method equally applies to  fermionic and bosonic Hubbard models. 
For concreteness,  we restrict our numerical studies to the case
of the Bose-Hubbard (BH) model, and term the resulting effective model the
``kinetically-driven Bose-Hubbard" (KDBH) model. We explicitly validate the relation of the KDBH model to the bosonic SYK model by comparing their
spectral form functions and out-of-time ordered correlation functions (OTOCs), using exact diagonalization methods. In this way, we
also investigate the behavior of the bosonic SYK model, which in contrast to the fermionic case, has remained largely unexplored so far~\cite{georges2000mean,georges2001quantum,fu_sachdev,baldwin2020quenched,swingle2024bosonic,liu2025dissipative}. Our proposed scheme offers a practical route for the experimental exploration of SYK-type models in optical-lattice setups, where kinetic driving can be implemented by simply shaking the lattice~\cite{eckardt2017colloquium,weitenberg2021tailoring,kinetic_epl}.

\textit{Model ---}
{We begin with the Bose-Hubbard (BH) model~\cite{jaksch1998cold,greiner2002quantum} describing bosonic particles, moving on a 1D lattice
\begin{equation}
H = -J \sum_j \left( b_j b_{j+1}^\dagger + \mathrm{H.c.} \right) +
U \sum_j b_j^\dagger b_j^\dagger b_j b_j \ ,
\label{hubbard}
\end{equation}
where $b_j / b^\dagger_j$ are bosonic annihilation / creation
operators acting on lattice site $j$. 
The model includes two processes: nearest-neighbor hopping
parameterized by the hopping amplitude $J$, and the onsite Hubbard repulsion $U$; see the sketch in Fig.~\ref{schematic_plot}(a).
For periodic boundary conditions, $H$ can be represented graphically as shown
in Fig. \ref{schematic_plot}(c), where each site is simply linked by $J$ to its nearest neighbors. We emphasise that the approach described below can also be readily applied to spinless fermions, in which case the bare interaction terms act between nearest-neighbor sites; see End Matter for a study of the fermionic case.}

Following Ref. \cite{kinetic_epl}, we now consider shaking the lattice such that
$J\!\rightarrow J(t)$ varies periodically with time. Importantly, we require that its time-average is zero over a driving period. By making
the particular choice $J(t)\!=\!J_0 \cos \omega t$,
one can obtain an expression for the
effective Hamiltonian in the high-frequency limit $\omega\!\gg\!U$, corresponding to the lowest-order term in the Magnus expansion~\cite{magnus,goldman2014periodically,bukov2015universal,eckardt2017colloquium}.
For the Bose-Hubbard model in Eq.~\eqref{hubbard},
the effective Hamiltonian takes the form
\begin{equation}
\Hkdbh = U \sum_{i,j,k,l} \ Q_{i j k l} \ b^\dagger_i b^\dagger_j b_k b_l \ ,
\label{effective_H}
\end{equation}
where the interaction amplitudes $Q_{i j k l}$ depend on the driving parameters; see the End Matter for a derivation of the effective Hamiltonian, in both the bosonic and fermionic cases, and for a discussion on the validity of this effective description.
Importantly, the single-particle hopping processes present in
the undriven Hubbard Hamiltonian $H$ have been eliminated by the drive, as required for realizing a SYK-type model. In fact, since the hopping energy scale $J$ has disappeared from the problem, the bare interaction strength $U$ now simply enters
as an overall energy scale. Henceforth, without loss of generality,
we set $U\!=\!1$. We point out that single-particle processes are effectively present in the micromotion, but they do not influence the full dynamics at stroboscopic times, as we further analyze in the End Matter.

We now compare the obtained KDBH Hamiltonian with the target (bosonic) SYK model. We restrict
ourselves to considering the real variant of the model,
\begin{equation}
\Hsyk = \sum_{i, j, k, l} \ J_{i j; k l} \ b^\dagger_i b^\dagger_j b_k b_l \ ,
\label{H_SYK}
\end{equation}
where $J_{i j; k l}$ are real random variables sampled from a Gaussian distribution with zero
mean and unit variance, and which satisfy bosonic exchange statistics and Hermiticity,
\begin{equation}
J_{i j; k l} =  J_{j i; k l} =  J_{i j; l k} =  J^\ast_{k l; i j} \ .
\label{conditions}
\end{equation}
As shown in Fig.~\ref{schematic_plot}(d), these interaction terms link every site to
every other site, the amplitude of each link being a random number.

Examining $\Hkdbh$ immediately reveals that it has the same structure of 4-operator terms as
$\Hsyk$, and that the conditions (\ref{conditions}) are automatically satisfied by
the amplitudes $Q_{i j k l}$. The two Hamiltonians only differ in that the $Q_{i j k l}$
are not random, but depend in a somewhat complicated way on the ratio of driving parameters,
$\kappa = J_0 / \omega$; see End Matter.
Although in principle these amplitudes can be long-ranged, in practice they decay approximately exponentially~\cite{pieplow_thesis} as the separation between the sites increases. This produces the
situation depicted in Fig. \ref{schematic_plot}(b), which is intermediate between the BH and
the SYK cases. Clearly the connectivity of the lattice is much higher than the BH case, but the
smallness or absence of some of the $Q_{i j k l}$ amplitudes means that the $\Hkdbh$ is considerably
more sparse than $\Hsyk$.

Sparse versions of the SYK model have been studied previously.
Early work on small systems \cite{cao_simulation, preskill} indicated that they reproduced SYK physics as long as more than 10\% of the matrix elements were non-zero. This estimate was later refined \cite{garcia_garcia_2021} to requiring at least $k N$ of the $N^4$ matrix elements to be non-zero, with $k \gg 1$.
Symmetry under the transformation $k \rightarrow \pi - k$ means that $\Hkdbh$ has a sparsity of 50 \%
\footnote{When the number of lattice sites is even, this symmetry of the KDBH Hamiltonian means that exactly half of the amplitudes will be zero, namely those $Q_{i j k l}$ such that $\left( i + j + k + l \right)$ is odd. For an odd number of sites, the cancellation is no longer exact, but nonetheless the amplitude of these terms is extremely small ($< 10^{-7}$ for the system sizes we consider).},
independent of the system size, so this condition is amply  fulfilled in our case.
As illustrated in Fig.~\ref{schematic_plot}(b), we find that the distribution of the amplitudes differs
from the dense Gaussian distribution of the SYK model. We will see, however, that
these twin effects, the sparsity of the Hamiltonian and the non-Gaussian distribution of the amplitudes,
do not prevent $\Hkdbh$ from emulating the SYK model with excellent fidelity.

\textit{Results ---} 
To investigate how closely the KDBH model reproduces
SYK physics, we will first consider its chaoticity.
A commonly used probe of quantum chaos is the distribution of
spectral spacings. According to the Berry-Bohigas conjecture, integrable systems will exhibit 
a Poissonian distribution of spacings, while chaotic systems will instead be described by
random matrix theory. An extensive
study of the spectral statistics of the KDBH model was performed in Ref. \cite{spectral_statistics}, demonstrating
that for $\kappa > 0.4$ it followed the Gaussian orthogonal ensemble~\footnote{When the number of sites and the number of particles are both even, a hidden symmetry
splits the distribution into being the sum of two GOEs. In this work we choose both parameters
to be odd to consider just the single GOE case.},
the same distribution as the bosonic SYK model~\cite{fu_sachdev}.

The spectral distribution measures quantum chaos at timescales longer than $1/\delta$, 
where $\delta$
is the mean energy spacing. To make a complete comparison between the models,
we must also consider their correspondence at short to intermediate times
by calculating the spectral form factor (SFF), defined as
$\SFF(t) = \left|  \sum_j \exp \left( i E_j t \right) \right|^2 / D^2$, where $D$ is
the dimension of the Hilbert space, and $E_j$ are the energies
of the Hamiltonian (\ref{effective_H}). In order to extract the spectral
distribution correctly, it is important to remove accidental symmetries, and block diagonalize 
the Hamiltonian in a particular symmetry sector~\cite{spectral_statistics}.
We follow the same procedure here when calculating the SFF,
and work in the zero momentum, reflection-symmetric subspace.

\begin{figure}
\begin{center}
\includegraphics[width=0.5\textwidth,clip=true]{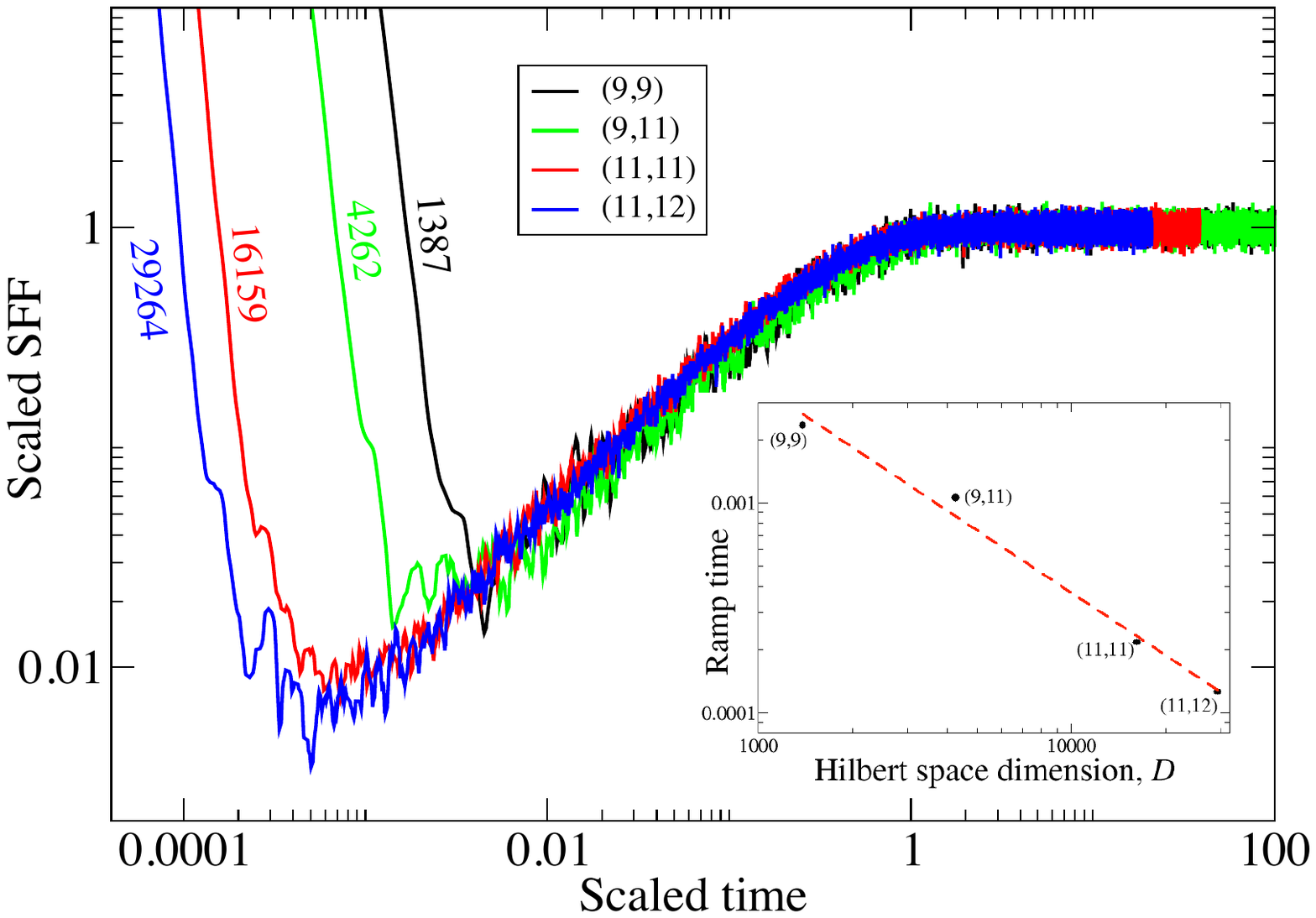}
\end{center}
\caption{Spectral form function (SFF) of the KDBH model for four different lattice sizes and boson numbers;
the notation $(N,L)$ denotes $N$ bosons on an $L$-site lattice. Multiplying the SFF by the
dimension of the Hilbert space $D$, and scaling the time by $1/D$ collapses the curves
to a universal form -- an initial drop, then a linear ramp, followed by a constant plateau.
Each curve is labeled by the corresponding values of $D$, and we see the ramp behavior
occurring earlier as $D$ increases.
Inset - We define the ramp time as the time at which the scaled SFF falls to
a value of 0.1. The red-dashed line drops as $D^{-1}$, and is included as a guide to the eye.}
\label{sff_plot}
\end{figure}

For many-body chaotic systems the SFF exhibits a particular structure: a non-universal
drop at the Thouless time, followed by a linear ramp, which then transforms into a flat
plateau $\SFF(t) \rightarrow 1/D$ for late times, $t > 1/\delta$.
In Fig. \ref{sff_plot}, we show the SFF obtained for various combinations of 
$(N,L)$, where $N$ denotes the number of bosons in the system, and $L$ is the number of sites.
We observed that the form of the SFF did not depend on the value of $\kappa$, and
so to reduce the size of fluctuations, we display the average of four different values of $\kappa$
($\kappa = 0.6, 0.7, 0.8, 0.9$).
Each value of $\kappa$ corresponds to a distinct set of $Q_{i j k l}$. This averaging procedure thus allows us to compare with the results
of the SYK model, which are obtained as an average over a Gaussian distribution of the hopping coefficients $J_{i j; k l}$.
The data presented were further smoothened by computing a running 50-point
average of the raw data. Scaling the SFF by a factor of $D$ and the time coordinate by $1/D$
collapses the curves onto each other,
displaying the required universal drop-ramp-plateau structure. From the graph it is clear that
the Thouless time reduces as $D$ increases, but the large non-universal fluctuations in
the SFF in the vicinity of the transition to the ramp
make it hard to measure this accurately. As a proxy, we define the {\em ramp time} as
the time at which the scaled SFF first drops to a value of 0.1 of the plateau value,
and plot this in the inset. This reveals that
the ramp time scales quite accurately as $1/D$.

Note that reproducing the universal form of the SFF is a stricter condition than demonstrating
the presence of quantum chaos via the spectral statistics. The Bose-Hubbard model,
for example, displays GOE statistics for $U/J < 0.1$ \cite{kollath}, but its SFF does not exhibit
the drop-ramp-plateau structure; see End Matter. 
This criterion was also used in Ref.~\cite{preskill} to determine
when sparse SYK models failed to adequately reproduce SYK physics, the absence of the ramp corresponding
to the loss of spectral rigidity and the transition to non-holographic behavior.

As a final point of comparison between the two models, we consider the out-of-time
ordered correlation function (OTOC). The decay of this quantity measures the degree
of ``scrambling'' in the system, that is, the delocalization of initially localized quantum information,
and in particular the OTOC reveals the presence of quantum chaos at short timescales.
The OTOC is a four-point correlation function of two local observables, and it is convenient
to use a normalized version \cite{zoller},
\begin{equation}
O(t) = \frac{\mathrm{Tr}\left[\rho \ W^\dagger(t) V^\dagger W(t) V \right]}  
{\mathrm{Tr} \left[\rho \ W(t)^2 V^\dagger V \right]} \ .
\label{otoc}
\end{equation}
We take the system's temperature to be infinite ($\beta \rightarrow 0$) so that $\rho = I$ and all
states enter the trace with equal weight. We have considerable freedom in our choice of
the local operators $V$ and $W$, as the scrambling dynamics should have a universal form,
and we select the unitary operators  $V = \exp \left[ i \pi n_1 \right]$ and $W = \exp \left[i \pi n_j
\right]$, where $n_j$ is the number operator acting on lattice site $j$. 
Since $V$ and $W$ take values of $\pm 1$ when operating on a number state, they are a natural
generalization of operators based on the fermionic number operator.

\begin{figure}
\begin{center}
\includegraphics[width=0.5\textwidth,clip=true]{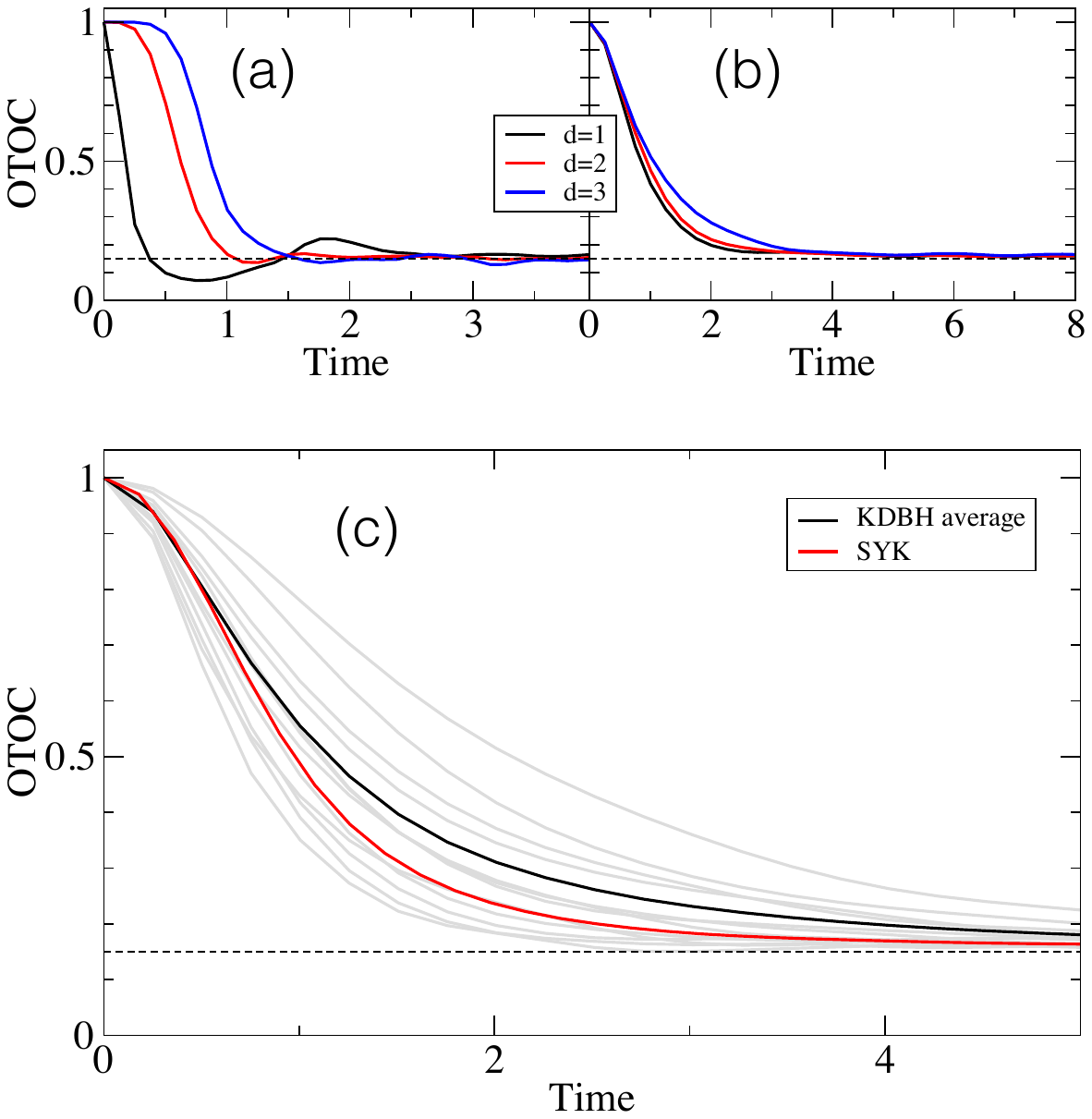}
\end{center}
\caption{(a) OTOCs for the BH model for 6 particles on a 6-site lattice with periodic
boundary conditions. When the operators are separated by a single lattice spacing ($d=1$)
the OTOC rapidly decays to its asymptotic value. For larger spacings, $d=2$ and $d=3$ there
is a delay until the OTOC begins its decay, corresponding to information spreading through
the lattice at a finite speed.
(b) As in (a) but for the KDBH model, with $\kappa = 0.8$. The OTOCs show no
sign of a time delay, beginning their decay immediately. This indicates the lack
of locality in this model.
(c) Gray curves indicate the OTOCs obtained for the KDBH model for
$\kappa=0.6, 0.7, 0.8, 0.9$ for the three different separations. Their average,
shown in black has the same form as the SYK OTOC, averaged over 25 random samples of $J_{i j;k l}$. 
To obtain quantitative agreement, the time coordinate of SYK data 
has been scaled by a factor of 9 (see text).}
\label{otoc_plot}
\end{figure}

In Fig.~\ref{otoc_plot}(a), we show the OTOCs obtained for the BH model ($U = J$) for the three different
spacings available: $d=1$ (nearest-neighbor, i.e.~$W$ is evaluated at lattice site $j\!=\!2$), $d=2$ (next-nearest neighbor, $j=3$) and
the largest separation $d\!=\!3$; these calculations are performed on a 6-site lattice with periodic boundary conditions. We can see that the $d=1$ result rapidly decays to 
a small but non-zero value. However, for
larger separations ($d\!=\!2,3$) there is a clear time delay before the decay occurs. This corresponds to
information propagating through space at a finite rate, given by the butterfly velocity. In contrast
to the SYK model, the BH chain thus exhibits spatial locality: an excitation made at a particular
point propagates outwards in a light-cone structure.

This is very different to the behavior exhibited by the KDBH model, as shown in Fig.~\ref{otoc_plot}(b) 
for $\kappa\!=\!0.8$.
The three curves show a very similar decay, and they are no longer ordered by their value
of separation $d$. This corroborates our expectation that the lattice connectivity seen in 
Fig. \ref{schematic_plot}(b) substantially reduces the degree of
locality from the BH case, due to the presence of effective long-range tunneling.
The decay rate shows little dependence on the amplitude of the driving, and in Fig.~\ref{otoc_plot}(c)
we plot the full set of OTOCs obtained for four values of $\kappa$. Averaging these 12 curves gives
a result which agrees strikingly with that of the bosonic SYK model.  
The quantitative comparison depends on just one fitting parameter, a rescaling of time in the
SYK model corresponding to the different energy scales of the two models
arising from the distinct distributions of hopping amplitudes -- the SYK model
has a Gaussian distribution with unit variance, while the KDBH model
has a narrower non-Gaussian form.

\textit{Concluding remarks.---}
We summarize our comparative studies of the (bosonic) SYK, Bose-Hubbard and KDBH models in Table~\ref{table_summary}. These results validate that SYK physics can be effectively realized using a practical periodic driving scheme.

\begin{table}
\begin{center}
\begin{tabular}{ |l|c|c|c| }
 \hline
{ } & SYK & Bose-Hubbard & KDBH \\
\hline \hline
Long-range hopping & Yes & No & Yes \\
RMT statistics & Yes & Yes & Yes  \\
Universal SFF & Yes & No & Yes \\
OTOC decay & Yes & No & Yes \\
 \hline
\end{tabular}
\caption{Comparison of the considered models. The Bose-Hubbard model exhibits chaos in its spectral statistics, but otherwise does not exhibit any of the features of the SYK model. In contrast, the KDBH model indeed reproduces all of the four criteria commonly used to identify SYK physics.}
\label{table_summary}
\end{center}
\end{table}

As previously noted, the KDBH model can be implemented by modulating the Hubbard hopping amplitude $J$ periodically in time, such that the average $\langle J(t)\rangle\!=\!0$ over each driving period~\cite{kinetic_epl}. In practice, this can be achieved through a two-frequency drive:~a lattice shaking of strength $A$ and frequency $\omega_1$ would renormalize the hopping strength according to a Bessel function $J_{{\rm eff}}\!=\!\mathcal{J}_0 (A/\omega_1)$~\cite{tien,arimondo,eckardt2017colloquium}; a modulation of the strength $A(t)$, at frequency $\omega_2\!\ll\omega_1$, and in the vicinity of the Bessel-function root, would then realize the desired hopping modulation. This approach assumes a proper separation of energy scales, which is crucial in view of limiting heating, e.g. through inter-band processes~\cite{eckardt2017colloquium}. Another promising direction relies on synthetic dimensions~\cite{celi2014synthetic}. In this framework, one would induce and modulate couplings between a set of atomic internal states spanning a fictitious lattice. Finite-range interactions would be required along the synthetic dimension, which can be achieved through the scheme proposed in Ref.~\cite{barbiero2020bose}.  

An interesting question concerns the experimental detection of SYK signatures in a Hubbard quantum simulator. An appealing approach would be to extract OTOCS by monitoring statistical correlations between randomized measurements~\cite{zoller} or through digital interferometric approaches~\cite{google2025observation}. Universal spectral signatures could be accessed by spectroscopy~\cite{roushan2017spectroscopic}. In principle, the long-range hopping amplitudes and fast scrambling could also be revealed by studying the dynamics of a mobile impurity immersed into the medium.

It would be worthwhile to investigate the potential effects of engineered disorder within our framework. As we show in the End Matter, breaking the translational symmetry of the KDBH model with a weak link provides a substantial improvement in the statistical distribution of the hopping amplitudes, and introducing other forms of disorder could also yield similar benefits. Finally, extending the framework to higher-dimensional systems should be straightforward and could open up intriguing avenues for exploration.

\begin{acknowledgments}
\textbf{\textit{Acknowledgments.---}} We thank Antoine Georges, Subir Sachdev, Lauriane Chomaz, Oriana Diessel, Peter Schlagheck and Sylvain Nascimbene for discussions. This research was financially supported by the ERC Grant LATIS, the EOS project CHEQS and the Fondation ULB. CEC and FS were supported by the Spanish MICINN through grant no. PID2022-139288NB-I00.
\end{acknowledgments}

\bibliography{syk_bib}

\appendix

\renewcommand{\theequation}{EM\arabic{equation}}
\setcounter{equation}{0}
\renewcommand{\thefigure}{EM\arabic{figure}}
\setcounter{figure}{0}

\begin{center}
{\bf End Matter}
\end{center}

{\em Derivation of the effective Hamiltonian for the driven Bose-Hubbard system -- }
We first consider the standard BH model (\ref{hubbard}), and allow the hopping to vary periodically in time, $J(t) = J_0 \cos \omega t$ 
\begin{equation}
H = -J_0 \cos \omega t \ \sum_j \left( b_{j+1}^\dagger b_j + \mathrm{H.c.} \right) +
U \sum_j b_j^\dagger b_j^\dagger b_j b_j \ ,
\label{single_species}
\end{equation}
where $j$ labels the lattice sites, and we take periodic boundary conditions.
We now Fourier transform this
Hamiltonian to momentum space to obtain
\begin{eqnarray}
H(t) &=& -2 J_0 \cos \omega t \ \sum_{k} \cos k \ b^\dagger_{k} b_{k} \nonumber \\
&+& \frac{U}{2 L} \sum_{k_1, k_2, k_3, k_4} \delta_{k_1 + k_2 , k_3 + k_4}
b^\dagger_{k_4} b^\dagger_{k_3} b_{k_2} b_{k_1} \ ,\notag
\end{eqnarray}
where the $\delta$-function enforces quasimomentum conservation, and the 
summations are taken over the first Brillouin zone.
Clearly when time-averaged over one period of the driving, the first term
in this equation vanishes, meaning that, as in the SYK model itself,
single-particle hopping terms are absent. 

To ensure convergence of the high-frequency expansion~\cite{goldman2014periodically,goldman2015periodically,bukov2015universal}, we now transform to the interaction picture, $H'(t) = i {\dot W}(t) W^\dagger(t)
+W(t) H(t) W^\dagger (t)$, where the unitary operator $W(t)$ is
defined as,
\begin{equation}
W(t) = \exp \left[ -2 i \kappa
\sin \omega t \ \sum_{k} \cos k \ b_{k}^\dagger b_{k} \right] \ ,\label{eq_change_basis_micro_motion}
\end{equation}
and $\kappa = J_0 / \omega$.
To perform this transformation it is advantageous to define
the operator 
\begin{eqnarray}
f(\lambda) &=& e^{-i \lambda \cos k \ b^\dagger_k b_k} 
b^\dagger_{k_4} b^\dagger_{k_3} b_{k_2} b_{k_1}
e^{i \lambda \cos k \ b^\dagger_k b_k}  \nonumber \\
{ } &=& U(\lambda) b^\dagger_{k_4} b^\dagger_{k_3} b_{k_2} b_{k_1} U^\dagger(\lambda) \ .
\label{f_operator}
\end{eqnarray}
In order to obtain $f(\lambda)$ in closed form, we first differentiate it with respect to $\lambda$,
\begin{equation}
\partial_\lambda f(\lambda) = i \cos k U(\lambda)
\left[ b^\dagger_{k_4} b^\dagger_{k_3} b_{k_2} b_{k_1}, b^\dagger_k b_k \right] U^\dagger (\lambda) \ .
\label{derivative}
\end{equation}
As shown in Ref. \cite{kinetic_epl}, this differential equation has the solution
\begin{equation}
f(\lambda) = e^{i \lambda \cos k \left(\delta_{k k_1} + \delta_{k k_2}
- \delta_{k k_3} - \delta_{k k_4} \right)} b^\dagger_{k_4} b^\dagger_{k_3} b_{k_2} b_{k_1} \ ,
\end{equation}
which enables the transformed Hamiltonian to be expressed as
\begin{widetext}
\begin{equation} 
H'(t) = \frac{U}{2 L} \sum_{k_1, k_2, k_3, k_4}
\delta_{k_1 + k_2, k_3 + k_4} e^{2 i \kappa \sin \omega t
\left( \cos k_1 + \cos k_2  - \cos k_3 - \cos k_4 \right)}
b^\dagger_{k_4} b^\dagger_{k_3} b_{k_2} b_{k_1} \ .\label{eq_this_result}
\end{equation}

Considering the high-frequency regime $\omega\!\gg\!U$, we calculate the lowest-order term in the Magnus expansion~\cite{magnus,goldman2014periodically,bukov2015universal,eckardt2017colloquium} by time-averaging Eq.~\eqref{eq_this_result} over one driving period. Making use of the Jacobi-Anger
expansion, this gives the effective Hamiltonian
\begin{equation}
\Heff = \frac{1}{T} \int_0^T dt H'(t) = \frac{U}{2 L} \sum_{k_1, k_2, k_3, k_4}
\delta_{k_1 + k_2, k_3 + k_4} {\cal J}_0 \left(2 \kappa \left[
\cos k_1 + \cos k_2 - \cos k_3 - \cos k_4 \right] \right) 
\ b^\dagger_{k_4} b^\dagger_{k_3} b_{k_2} b_{k_1} \ ,
\end{equation}
where ${\cal J}_0$ is the Bessel function
of zeroth order. Transforming this result back to real space
yields the KDBH model,
\begin{equation}
\Hkdbh = U \sum_{i, j, k, l} Q_{i j k l} b^\dagger_i b^\dagger_j b_k b_l
\end{equation}
where the hopping amplitudes are given by, 
\begin{equation}
Q_{i j k l} = \frac{1}{2 L^3} \sum_{k_1, k_2, k_4} e^{i \left( k_1 (j - l) + k_2 (j - k) + k_4 (i - j) \right)}
{\cal J}_0 \left[ 2 \kappa \left( \cos k_1 + \cos k_2 - \cos \left( k_1 + k_2 - k_4 \right)
-\cos k_4 \right) \right] \ .
\label{onsite}
\end{equation}
\end{widetext}

We note that higher-order terms in the effective Hamiltonian would correspond to many-body interactions (beyond two-body interactions). These are neglected in the present study, which assumes a high frequency regime $\omega\!\gg\!U$. We have explicitly verified that
in the high-frequency limit
the quasienergies obtained from Eq.~\eqref{single_species} indeed reproduce the energy levels of $\Hkdbh$~\cite{spectral_statistics}. 

\begin{figure}
\begin{center}
\includegraphics[width=0.5\textwidth,clip=true]{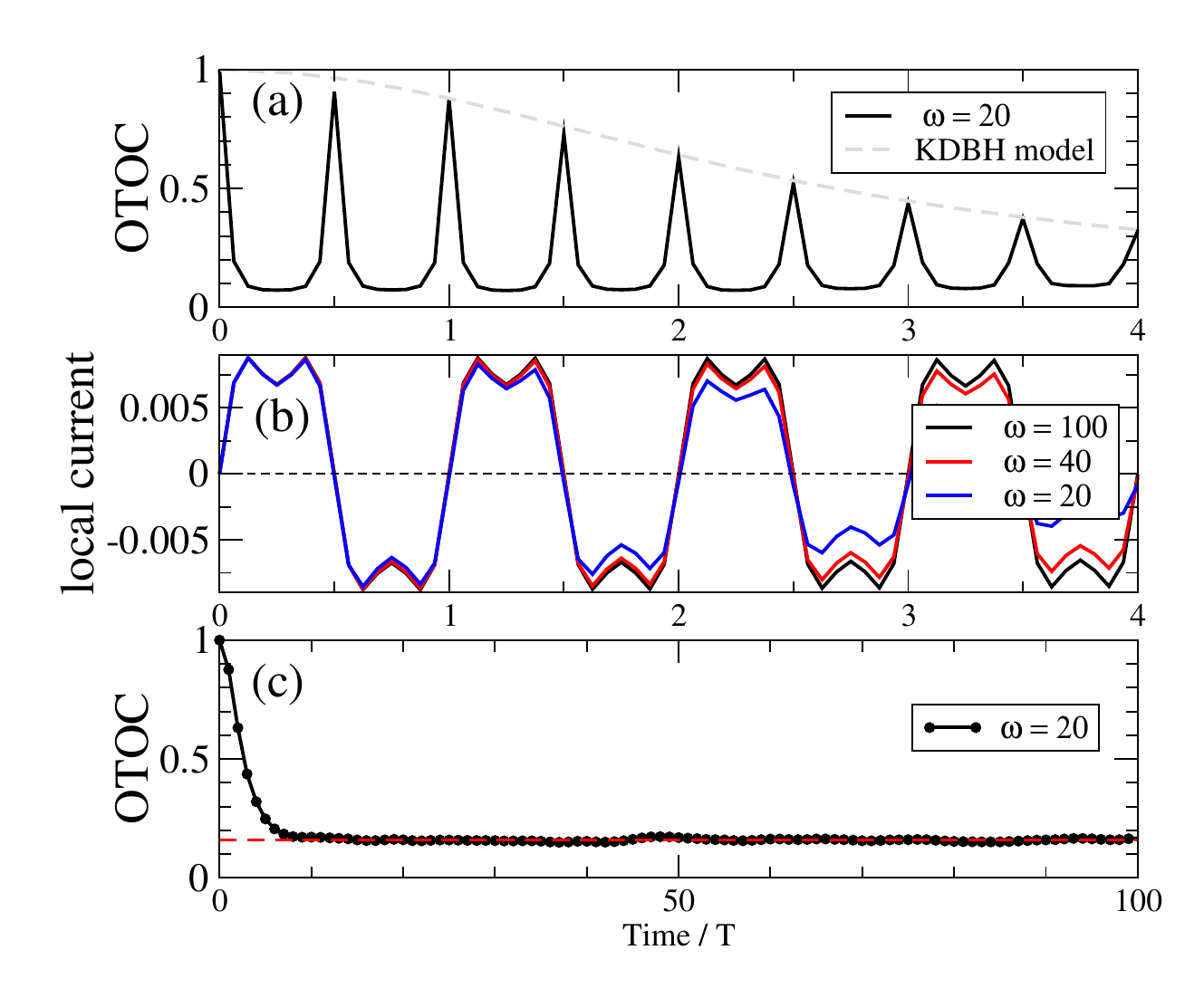}
\end{center}
\caption{(a) Comparison of the OTOC obtained from the KDBH model
with the full time-dependent dynamics of Eq.~\ref{single_species} with $\omega\!=\!20$. The stroboscopic values of the time-dependent model (at integer multiples of the driving period $T$) coincide
perfectly with the effective model; in between the micromotion
produces periodic deviations. (b) The local current, $2 \mathrm{Im} \langle b_2^\dagger b_3 \rangle$, measuring the
particle transfer between lattice sites $2$ and $3$ in the
full time-dependent model. As expected from the form of the micromotion operator [Eq.~\eqref{micro}], the amplitude of the current shows little dependence on $\omega$, at fixed $\kappa$. (c) Over many driving periods, the OTOC decays to a constant asymptotic value. The driven system shows no sign of heating, even over this long time-scale. The red-dashed line is included as a guide to the eye.
In all cases $\kappa\!=\!0.8$ and a 6-site lattice was used.} 
\label{heating_fig} 
\end{figure}

Furthermore, as we show in Fig.~\ref{heating_fig}(a), we verified that the
OTOCS in Eq.~\eqref{otoc} calculated from $\Hkdbh$ 
coincide with those calculated using the time-dependent Hamiltonian in Eq.~\eqref{single_species} at
stroboscopic sampling times, to an excellent degree of
accuracy. Importantly, in Fig.~\ref{heating_fig}(c) we also present the OTOC evaluated with the time-dependent Hamiltonian over a long timescale, 100 periods of driving at
a frequency of $\omega\!=\!20$. The OTOC initially decays rapidly,
as we have seen previously, and then tends towards a constant asymptotic value. That the OTOC behaves almost identically in the effective and full-time dynamics, both at short and longer times, clearly indicates that heating -- generically present in periodically-driven systems~\cite{eckardt2017colloquium} -- does not manifest over the time scales required to probe out-of-equilibrium SYK physics, in the high-frequency regime of our driving scheme.

\begin{figure}
\begin{center}
\includegraphics[width=0.5\textwidth,clip=true]{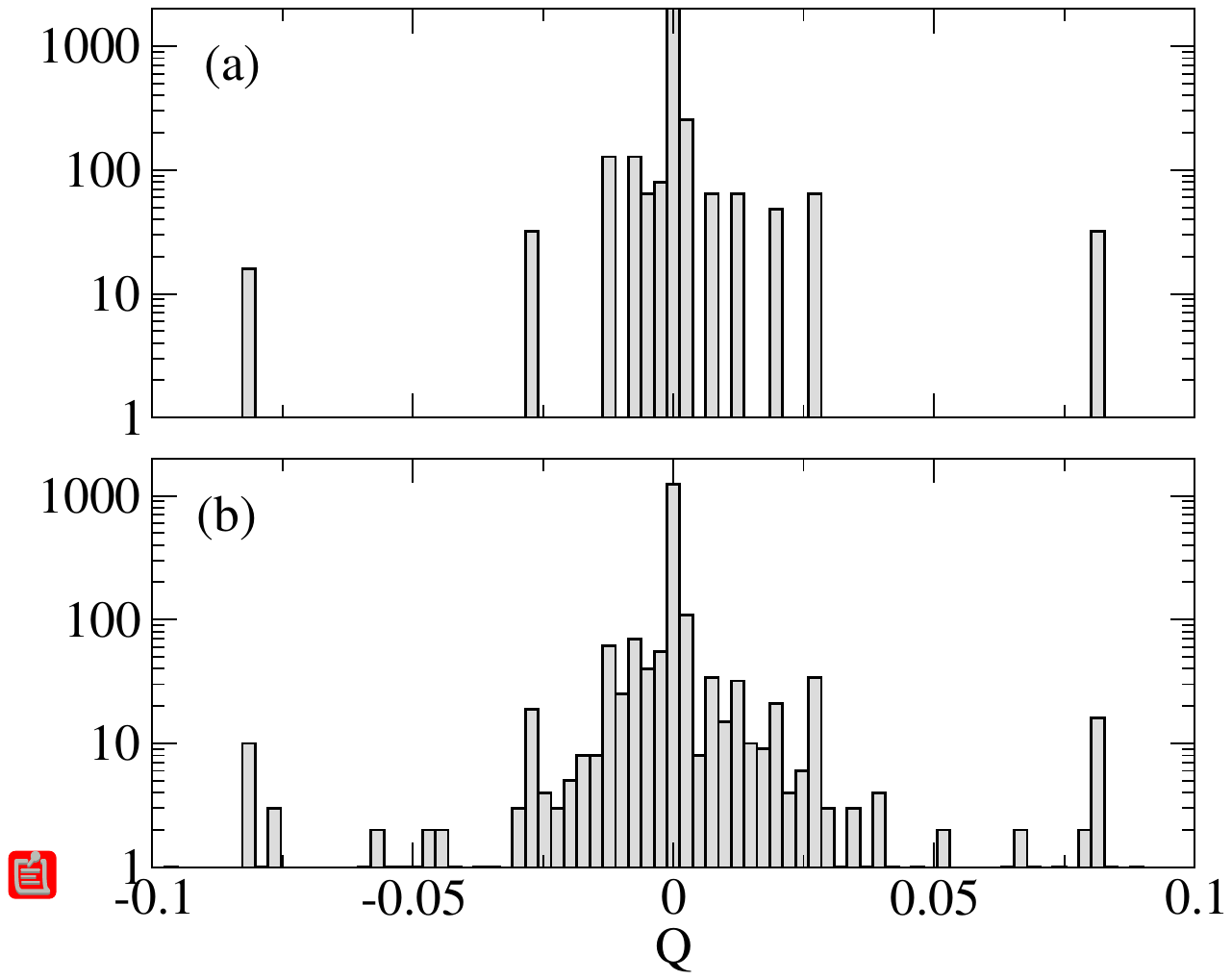}
\end{center}
\caption{Distributions of the $Q$ amplitudes in an 8-site KDBH model
for $\kappa = 0.6$. (a) Clean system: as seen previously, the distribution is
non-Gaussian and contains several gaps. (b) Introducing a weak link in the
parent BH Hamiltonian with $J = \left( J - \epsilon \right)$  produces a
distribution much closer to the target Gaussian form.}
\label{dists}
\end{figure}

\bigskip
{\em Breaking translational symmetry -- } As can be seen in Fig.~\ref{schematic_plot}(b), the distribution of the $Q_{i j k l}$
is far from Gaussian, and also exhibits various gaps. We attribute this feature to the translational invariance of the model, which leads to the spatial structures displayed in Eq.~\eqref{onsite}. To explore this effect, we now introduce a small defect that explicitly breaks translational symmetry:~following Ref.~\cite{kinetic_driving_2023}, we obtain the $Q$ amplitudes when one of the hoppings in the BH model differs from the others by $\epsilon$. We show the result in Fig.~\ref{dists}
for $\epsilon\!=\!0.001$. We find that this slight breaking of translational invariance is enough to refine the distribution of the $Q$ amplitudes, bringing it into closer alignment with the Gaussian form characteristic of the SYK model.

\bigskip
{\em Micromotion effects} -- The effective-Hamiltonian description considered in this work assumes that the dynamics is well captured by the stroboscopic evolution generated by the time-evolution operator $\hat U (T)$ over a period $T\!=\!2 \pi/\omega$; see Refs.~\cite{goldman2014periodically,bukov2015universal}. However, it is relevant to analyze the effects of the micromotion, which can be potentially significant at intermediate times (i.e.~within each period of the drive). This can be studied by writing the time-evolution operator in the form~\cite{goldman2014periodically,goldman2015periodically}
\begin{equation}
\hat U(t; t_0) = e^{-i \hat{\mathcal{M}} (t)} e^{-i (t-t_0) \hat H_{{\rm eff}}} e^{i \hat{\mathcal{M}} (t_0)},
\end{equation}
where $\hat{\mathcal{M}} (t+T)\!=\!\hat{\mathcal{M}}(t)$ captures the micromotion. 

At lowest-order in the $(1/\omega)$ expansion -- namely, in the high-frequency limit considered in this work -- the micromotion operator can be identified with the operator $W (t)$ introduced in Eq.~\eqref{eq_change_basis_micro_motion}; see Ref.~\cite{goldman2015periodically}. Using the real-space representation, the micromotion operator thus reads
\begin{equation}
\hat{\mathcal{M}} (t) = \exp \left[ - i \kappa
\sin \omega t \ \sum_{j} (b_{j+1}^\dagger b_{j} + {\rm H.c.}) \right] \ .
\label{micro}
\end{equation}
Hence, in this high-frequency regime, the micromotion simply corresponds to a time-periodic activation of single-particle hopping, within each period of the drive. We note that the amplitude of this micromotion is set by the parameter $\kappa$.

Since SYK physics is captured by a purely interacting Hamiltonian (without single-particle processes) -- see Eqs.~\eqref{effective_H}-\eqref{H_SYK}  -- it is relevant to analyze how such micromotion affects the full-time dynamics of our driven system. To do so, in Fig.~\ref{heating_fig}(b) we plot the full-time dynamics, including the micromotion, of the local
current, $2 \mbox{Im} \langle b_j^\dagger b_{j+1} \rangle$, for various drive frequencies. Using this observable allows us
to track when single-particle hopping is active. In all cases, the local current oscillates with the same frequency as the driving, and at stroboscopic times it is equal to zero. As
described by Eq.~\eqref{micro}, the amplitude of the micromotion
depends solely on $\kappa$ in the high-frequency regime, and so does not alter much as $\omega$ is varied.

\bigskip
{\em Derivation of the effective Hamiltonian for a driven Fermi-Hubbard system -- }
The derivation for a fermionic system proceeds in much the same way as previously.
We must consider a different starting
Hamiltonian, however, as the Hubbard interaction written in Eq. \ref{hubbard} would identically vanish. Accordingly we introduce a Hubbard-type model with nearest-neighbor interactions
\begin{equation}
H = -J \sum_j \left( c_{j+1}^\dagger c_j + \mathrm{H.c.} \right) +
U \sum_j c_j^\dagger c_{j+1}^\dagger c_j c_{j+1} \ ,
\label{spinless}
\end{equation}
where $c_j / c^\dagger_j$ are spinless fermion operators.

Following the same procedure as before we obtain a very similar form for the $f(\lambda)$ operator, only differing from (\ref{f_operator}) by a momentum-dependent phase
\begin{equation}
f(\lambda) = e^{i (k_4 - k_2)} \ U(\lambda) c^\dagger_{k_4} c^\dagger_{k_3} c_{k_2} c_{k_1} U^\dagger(\lambda) \ .
\end{equation}
Noting that the commutator in Eq. \ref{derivative} does not depend upon
the particle statistics,
we can again obtain a closed form solution for $f(\lambda)$, and hence obtain the hopping amplitudes for this model
\begin{widetext}
\begin{equation}
Q_{i j k l} = \frac{1}{2 L^3} \sum_{k_1, k_2, k_4} e^{i \left( k_1 (j - l) + k_2 (j - k - 1) + k_4 (i - j + 1) \right)}
{\cal J}_0 \left[ 2 \kappa \left( \cos k_1 + \cos k_2 - \cos \left( k_1 + k_2 - k_4 \right)
-\cos k_4 \right) \right] \ .
\label{neighbour}
\end{equation}
\end{widetext}

Comparing the expressions for $Q_{i j k l}$ in the two models
reveals their strong similarity, and in fact (\ref{neighbour}) can be mapped
to (\ref{onsite}) by making the substitutions
$i \rightarrow i-1, \ k \rightarrow  k -1$. Consequently the distribution
of $Q_{i j k l}$ in the two models is identical, the difference being that
$Q_{i j k l}$ labels different processes in each case.

We emphasise that the amplitudes do not depend on the statistics of the particle -- the same result (\ref{neighbour}) is obtained for both bosons and spinless fermions. Accordingly we believe that kinetically-driven Hubbard models of both bosonic and fermionic type will permit the simulation of the corresponding SYK Hamiltonians.

\begin{figure}
\begin{center}
\includegraphics[width=0.5\textwidth,clip=true]{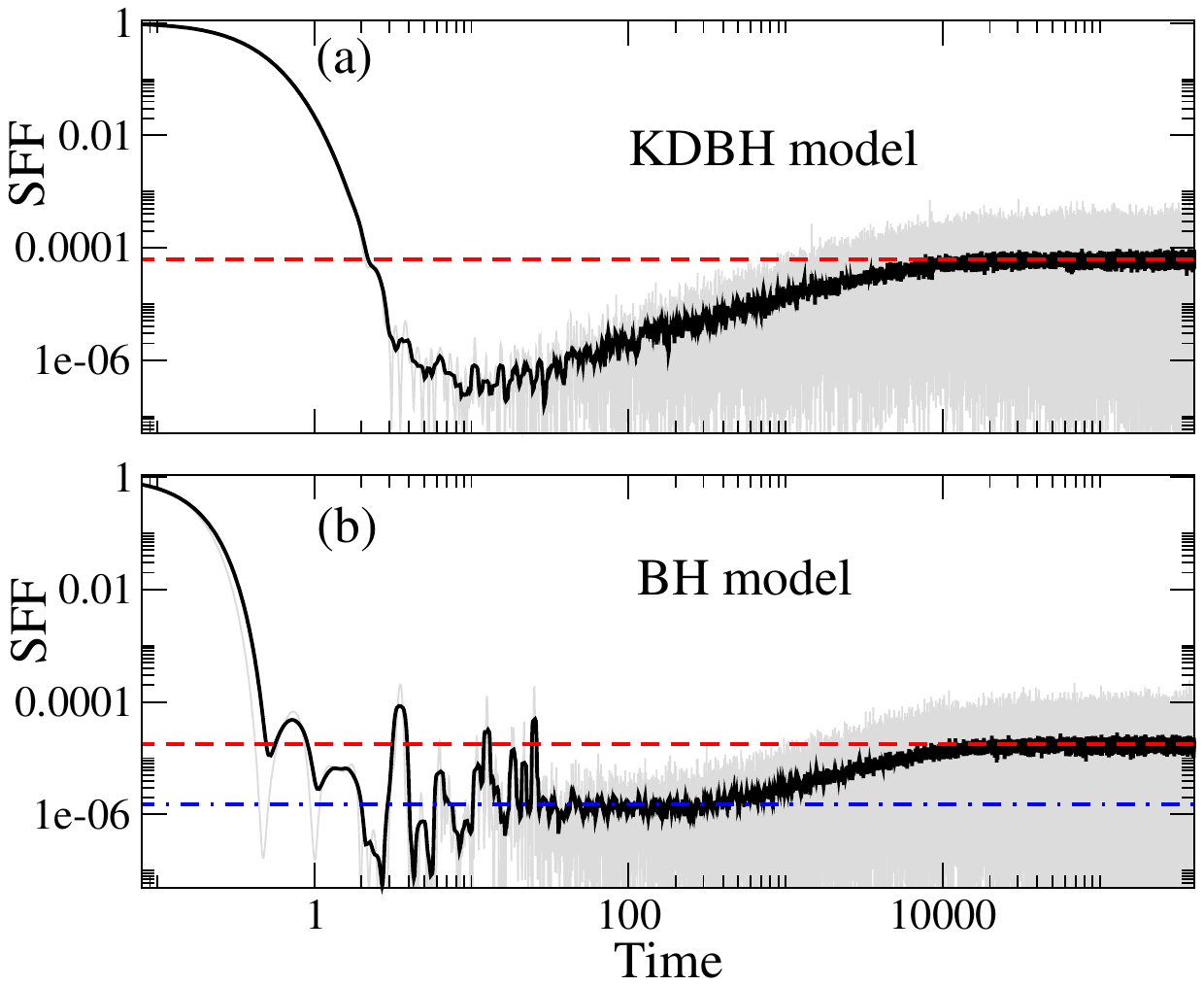}
\end{center}
\caption{(a) The SFF for the KDBH model ($\kappa = 0.8$ and $(N,L)=(11,11)$)
follows the
universal form: a sharp decay at the Thouless time, followed by a linear
ramp before finally reaching a plateau. The gray curve shows the
raw data, the black line is obtained by a 50-point running average to smoothen
the oscillations.
(b) In contrast, the SFF for the BH model ($U=0.2 J$ and $(12,12)$) does not have the universal form.
Following the initial decay, the SFF first oscillates
about a constant value, before going through the ramp-plateau
behavior at a much later time.
Red dashed lines indicate the value of $D^{-1}$ (where $D$ is the dimension of the
Hilbert space); the blue dot-dashed line is a guide to the eye indicating the first
plateau of the BH model.}
\label{app_figure}
\end{figure}

\bigskip
{\em SFF of the BH model -- }
The SFF of the KDBH model displays the universal form expected
of chaotic systems, consisting of a rapid drop up to the Thouless time,
followed by a linear ramp, and finally a plateau at long times. This
behavior can be clearly seen in Fig. \ref{app_figure}(a,) where we plot
a typical example,
the SFF for a $(11,11)$ system at $\kappa=0.8$.

As we note in the text and in Table \ref{table_summary}, the BH model does not follow this form.
In Fig. \ref{app_figure}(b) we show the SFF for a weakly interacting BH model,
which after the initial fall first goes through a preliminary plateau, before then
showing the ramp and final plateau at longer times. In contrast to the KDBH and SYK models, the
BH model thus only manifests chaos at long time scales, the initial plateau possibly
indicating a pre-thermalisation regime where the Hilbert space of the system
is not fully explored.

\end{document}